# Understanding Specimen- and Grain-Size Effects on Nanoscale Plastic Deformation Mechanisms and Mechanical Properties of Polycrystalline Yttria-Stabilized Tetragonal Zirconia Nanopillars

Ning Zhang and Mohsen Asle Zaeem[*1]

Department of Mechanical Engineering, Colorado School of Mines, Golden, CO 80401, USA

**Abstract**

Specimen- and grain-size effects on nanoscale plastic deformation mechanisms and mechanical properties of polycrystalline yttria-stabilized tetragonal zirconia (YSTZ) nanopillars are studied by molecular dynamics simulations. Through uniaxial compression of YSTZ columnar nanopillars, intergranular and transgranular deformation mechanisms are investigated. Cooperative intergranular deformations including grain boundary sliding and migration, grain rotation, and amorphous phase formation at grain boundaries are revealed. Results also reveal formation of partial dislocations, which act as splitters of large grains and play a significant role in facilitating the rotation of grains, and consequently promote amorphous-to-crystalline phase transition in-between neighboring grains. An increase in free surface-to-volume ratio is found to be responsible for specimen size-induced softening phenomenon, where a decrease in Young's modulus and strength is observed when the specimen width decreases from 30 nm to 10 nm. Also, a decrease in Young's modulus and strength is revealed with the decrease of average grain size from 15 nm to 5 nm. Grain boundary density is identified to be responsible for the observed grain size-induced softening behavior in polycrystalline YSTZ nanopillars. A transition in dominant deformation mechanism is observed from amorphous phase formation at grain boundaries to a competition between intergranular grain boundary sliding and transgranular phase transformation. Furthermore, an inverse Hall-Petch relationship is revealed describing the correlation between grain size and strength for polycrystalline YSTZ nanopillars with grain sizes below 15 nm.

**Keywords:** Zirconia nanopillar; Polycrystalline; Plastic deformation mechanism; Specimen size; Grain size.

---

[1]* Corresponding author. Email address: zaeem@mines.edu

## 1. Introduction

Polycrystalline yttria-stabilized tetragonal zirconia (YSTZ) ceramics are promising materials for some structural applications, because of their stress-induced tetragonal to monoclinic phase transformation at moderate temperatures [1, 2], which enables the design of self-healing zirconia-based shape memory materials. Also, their ferroelastic domain switching behavior at high temperatures makes them potentially suitable for high temperature actuation [3] and energy conversion [4] applications. Some of the macroscopic properties of polycrystalline YSTZ ceramics, such as their mechanical and electrical properties, have been reported to be closely related to their micro- and nano-structures [5-8], or their *intrinsic* characters. Since zirconia ceramics are most common in their polycrystalline form, formation and evolution process of defects (such as dislocation, crack, void, amorphous phase, etc.), as well as the interaction between defects and grain boundaries (GBs), are of particular interest. However, there is still uncertainty in fully understanding the GB effects on nanoscale deformation mechanisms and mechanical properties of YSTZ ceramics.

Among the *intrinsic* factors that govern the superplastic response of zirconia ceramics, chemical bonding state at GBs [9-11] has been reported repeatedly. In the last decade, most of the experimental research efforts have been focused on the intergranular fracture surfaces of polycrystalline YSTZ [12-14], which inevitably overlooked the evolution of GBs and transgranular deformations. Yttrium segregation to the GBs was found to affect the mechanical behavior of zirconia ceramics because of the formation of an electrically charged layer at the GBs [15]. Doped with a small amount of $Al_2O_3$, both transgranular phase transformation and grain-growth rate were enhanced [16], and also a GB segregation-induced phase transformation model was proposed [12]. Transgranular characteristic is also of particular interest, however only a few experimental and theoretical studies have focused on transgranular properties of polycrystalline YSTZ. For example, via theoretical analysis, transgranular dislocation movement and pile-ups were reported to play little or no significant role on the superplasticity behavior of polycrystalline YSTZ [17]. Kurumada et al. [18] investigated the assistance behavior of free mobile oxygen vacancies experimentally. Their results suggested that oxygen vacancies make a significant contribution to the transgranular electrical conduction of polycrystalline YSTZ. Very recently, the role of GBs in determining the strength and plastic deformation of YSTZ bicrystals was studied by molecular dynamics (MD) simulations [19]. It was found that the strength of YSTZ bicrystal changes when the loading direction switches from perpendicular to parallel to the GB orientation. Despite the previous studies on this topic, it is worth noting that a comprehensive understanding of the nanoscale deformation mechanisms of polycrystalline YSTZ, such as evolutions of GB mobility, grain interior deformation, dislocation motion and crack/void formation, are still lacking.

In addition to the nano/microstructural-dependent plasticity, size effects have emerged as another important factor in controlling the mechanical performance of materials, especially when the sample and/or microstructure size is reduced to nanometer scale. Indeed, in polycrystalline YSTZ ceramics, we can classify the size effect as *intrinsic*, i.e., grain size, when it is linked to the interior nanostructure, and *extrinsic* when it is related to the overall specimen size. Understanding and characterizing the mechanical response and the underlying deformation mechanisms of nanostructured shape memory ceramics are particularly important in design of reliable and effective shape memory devices.

It has now been ubiquitously demonstrated that the deformation behavior of materials with specimen dimensions down to nanoscale is very different from that of their corresponding bulk materials (micro and macro scale specimens), leading to specimen size-dependent strength [20, 21]. For example, it was shown for single crystal gold that its yield strength exhibits an order of magnitude increase over its bulk when the specimen size ranges from tens of nanometer to submicrometer scale [22]. In general, this "smaller is stronger" phenomenon can be interpreted by the dislocation nucleation-governed plasticity, which has been reported in a wide variety of metals (e.g., Ni, Au, Cu, Mo) [23-25] and ceramics [26]. On the other hand, in some semiconductor materials such "smaller is stronger" behavior can be induced by phase transformation [20, 21, 27]. For example, in a recent study of Si nanoparticles under compression, the hardness is found to increase when the diameter of particles decreases from 50 nm to 10 nm, and this was caused by the metastable phase transformation and consequent deformation twinning behavior during compression. On the contrary, a "smaller is weaker" trend was observed in single crystal 3 mol% YSTZ shape memory ceramic, whose plastic deformation is mediated by martensitic phase transformation [26]. As for the case of polycrystalline metals, some studies focusing on the effect of specimen size on strength of nanocrystalline fcc crystals have been conducted. For example, Jang and Greer [28] observed a "smaller is weaker" trend in a GB deformation-dominated Ni-4%W alloy with average grain size of 60 nm. Furthermore, some studies on nanocrystalline metals pillars [29, 30] and microwires [31] have manifested coupled effects of specimen size and grain size on strength; on micrometer size range an opposite trend, "smaller is stronger," by retarding dislocation motion was revealed in Ni [32] and Ag [31] microwires. Although specimen size effect on mechanical properties of polycrystalline metals is indeed a controversial topic, there is a lack of investigations on specimen size effect on mechanical properties and deformation mechanisms in shape memory ceramics.

Internal grain size is well known to have a significant effect on the mechanical behavior of nanocrystalline materials, in particular, on the yield stress. The dependence of yield stress on grain size in metals and alloys has been well established, i.e., via the Hall-Petch relation in polycrystals with ~40 nm and larger sized grains [33-35], and via the inverse Hall-Petch relation when the average grain size is reduced to below ~20 nm [36-38]. Such deviation of plastic mechanical response at different length scales is due to the engagement of different deformation mechanisms. In metals with an average grain size lager than ~40 nm, plastic deformation is governed by dislocations motion and their interaction with GBs, e.g., dislocations pile-up against GBs. However, as the average grain size decreases to below ~40 nm, multiple dislocations cannot be accommodated and alternative plastic deformation mechanisms such as GB sliding [39], partial dislocation emission and absorption at GBs [40], GB migration [41], and nanograin rotation [42] are expected. When the average grain size is reduced to below ~20 nm, GB-assisted deformation [42] is activated, which leads to softening via a well-known inverse Hall-Petch mechanism. As for the case of polycrystalline ceramics, most of the previous efforts have focused on the grain size effect on crystal structure [43, 44], phase transformation [43], dielectric and ferroelectric behaviors [45-47], and energy storage properties [48]. In particular, for zirconia ceramics, transformation toughening is strongly dependent on grain size. It was further found that nanocrystalline YSTZ with average grain size of ~85 nm has higher fracture toughness than submicrometer-grained zirconia ceramics [49], but it is difficult for experimental studies to provide precise description of toughening mechanisms.

Overall, despite the previous efforts that have been devoted to the study of zirconia ceramics, a comprehensive understanding of the specimen and grain size effects on mechanical properties of stabilized zirconia ceramic, as well as the corresponding fundamental deformation mechanisms at nanoscale, are lacking. Yet understanding the effects of these extrinsic (specimen size) and intrinsic (grain size) textures are likely to reveal new deformation mechanisms and significantly extend our knowledge of shape memory behavior in these ceramics.

In this regard, atomistic simulation is a very powerful tool to investigate the fundamental deformation mechanisms and structure-property relationship at the nanoscale. Indeed, MD simulations have been widely used to uncover the atomic-level details of interesting phenomena like phase transformation and dislocation migration [1, 20, 50], as well as GB structure [39, 40], and intergranular and transgranular plastic deformation mechanisms [51]. In this work, by utilizing MD simulations, we present results of polycrystalline YSTZ nanopillars under uniaxial compressive loading. The nanoscale deformation mechanisms, such as GB sliding, grain rotation, dislocation emission and absorption, and amorphous phase formation are revealed. We also observe global and local size-dependent softening behaviors, i.e., nanopillars with smaller specimen size or with smaller grain size have lower yield stresses. Specifically, we quantitatively examine the role of free surfaces and GB density in GB-assisted deformation, and construct inverse Hall-Petch equations for polycrystalline YSTZ nanopillars to describe the relation between average grain size and strength.

## 2. Simulation details and methodology

Polycrystalline YSTZ nanopillar structures are generated by using the Voronoi construction method [52] with random nucleated seeds and random crystallographic orientations. In this work, nanopillars with rectangular and square cross-sections are employed. The Voronoi tessellation is implemented by first picking the centers of all the grains randomly, and then letting atoms grow out from the center positions with the usual hexagonal crystal structure.

To clearly visualize and analyze the atomistic processes of GB sliding and defects formation, as well as the interaction between dislocation and GBs, a two-dimensional (2D) columnar nanopillar (grains are extruded in the y direction) is generated with a specimen size of 47.0 × 25.0 × 20.0 nm$^3$ and a total number of atoms of 1,926,000, as shown in Figure 1. This sample contains 10 grains with average grain size of 12.2 nm. Whereas for generality, in the studies of specimen size and grain size effects, fully three-dimensional (3D) polycrystalline YSTZ nanopillars are constructed by randomly selecting grain centers and their crystallographic orientations.

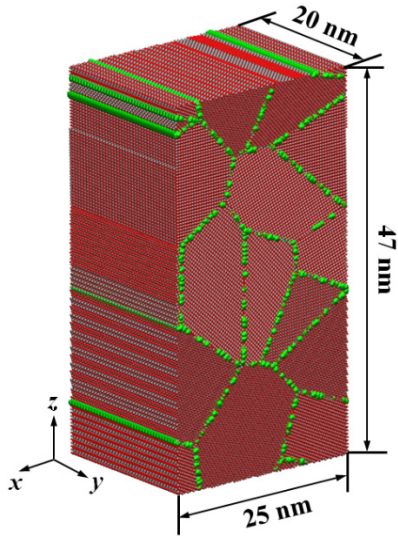

Figure 1. Computer model of a 2D columnar YSTZ nanopillar (grains are extruded in the y direction). The green atoms represent GBs.

In order to systematically reveal the specimen size effect and uncover the underlying deformation mechanisms in polycrystalline YSTZ nanopillars, five 3D models with square cross-sections, and the width (*w*) ranging from 10 nm to 30 nm, are considered, as shown in Table 1. These models are generated by first constructing the largest nanopillar model with the size of 30

× 30 × 40 nm³, and then cutting the smaller models from this large model. Therefore, all the obtained specimens have the same average grain size of 7.0 nm. For grain size effect study, six 3D polycrystalline YSTZ nanopillars with average grain sizes ranging from 5.0 to 15.0 nm are constructed, as summarized in Table 1. All the nanopillars have the same overall dimensions of 25 × 25 × 40 nm³. Table 1 summarizes the structural information of all related nanopillars in this study.

The LAMMPS code (Large-scale Atomic/Molecular Massively Parallel Simulator developed by Sandia National Laboratories) [53] is utilized in this work for conducting MD simulations. The empirical Born-Meyer-Buckingham (BMB) interatomic potential combine with long-range Coulomb term [1] is employed to describe the interactions between ions in polycrystalline YSTZ nanopillars. This interatomic potential had been shown to provide good agreements with experimentally measured properties, and also be capable of simulating the dislocation migration and martensitic phase transformation behaviors of single crystalline and bi-crystalline YSTZ [1, 19, 26]. The equations of motion are integrated with the velocity-Verlet method [54] with a time step of 1 fs for all simulations. To obtain equilibrated nanostructures, the initially generated nanopillars are fully relaxed at room temperature (298 K) for 0.1 ns using the constant volume and constant temperature (NVT) ensemble. Thereafter, uniaxial compressive loading (displacement controlled) is performed along $z$-direction (the long axis). The Nose-Hoover thermostat [55] is used to maintain the temperature at a constant value of 298 K throughout the simulations. A relatively slow strain rate of $1 \times 10^8$/s, which has been demonstrated to be slow enough in order to produce strain rate independent results by MD simulations [1, 20, 50], is applied. Coordinate number (CN) is adopted to identify the evolution of GB deformation, dislocation migration and phase transformation.

Table 1. Specimen size, average grain size (*d*), and the number of total atoms (*N*) of polycrystalline YSTZ nanopillars used in this study.

| Purpose of study | Dimension of nanopillars | Specimen size (nm$^3$) | *d* (nm) | *N* |
|---|---|---|---|---|
| Deformation mechanism | 2D columnar | 20.0 × 25.0 × 47.0 | 12.2 | 1,926,000 |
| Specimen size effect | 3D | 10.0 × 10.0 × 40.0 | 7.0 | 315,000 |
| | | 15.0 × 15.0 × 40.0 | | 710,000 |
| | | 20.0 × 20.0 × 40.0 | | 1,287,000 |
| | | 25.0 × 25.0 × 40.0 | | 1,971,000 |
| | | 30.0 × 30.0 × 40.0 | | 2,838,000 |
| Grain size effect | 3D | 25.0 × 25.0 × 40.0 | 5.4 | 1,938,000 |
| | | | 6.2 | 1,956,000 |
| | | | 7.8 | 1,983,000 |
| | | | 9.8 | 2,005,000 |
| | | | 12.4 | 2,025,000 |
| | | | 14.7 | 2,032,000 |

## 3. Simulation results and discussion

### *3.1 Plastic deformation mechanisms*

Figure 2 shows snapshots of the nanostructural evolution of 2D columnar polycrystalline YSTZ nanopillar under uniaxial compression. The onset of plastic deformation is manifested by deformation of atoms around GBs and intergranular GB sliding at GBs, which make approximately $45^0$ angle with the loading direction. After compression, some corrugations at the free surfaces along GBs are observed (Fig. 2b, c). In addition, GB sliding produces geometrical misfits at GBs and triple junctions, as indicated by thick orange arrows in Fig. 2a-c. With continuing friction in-between grains, voids on GBs are formed and stretched to grow, then coalescence to generate cracks (denoted by thick yellow arrows in Fig. 2a'-c'), which is one of the causes of intergranular fracture. It should be noted that GBs are defective regions containing pre-existing voids and misoriented atoms, thus, they have a lower load carrying capacity than the grain interiors. Therefore, the intergranular cracking is due to the fact that GBs are weaker than the crystalline grains. Meanwhile, along with the GB migration, the thickness of amorphous phase at GBs keeps increasing, which can be confirmed by the green atoms at the GBs. The formation of amorphous phase at GBs can be considered as one of the main superplastic deformation behaviors of polycrystalline YSTZ nanopillars.

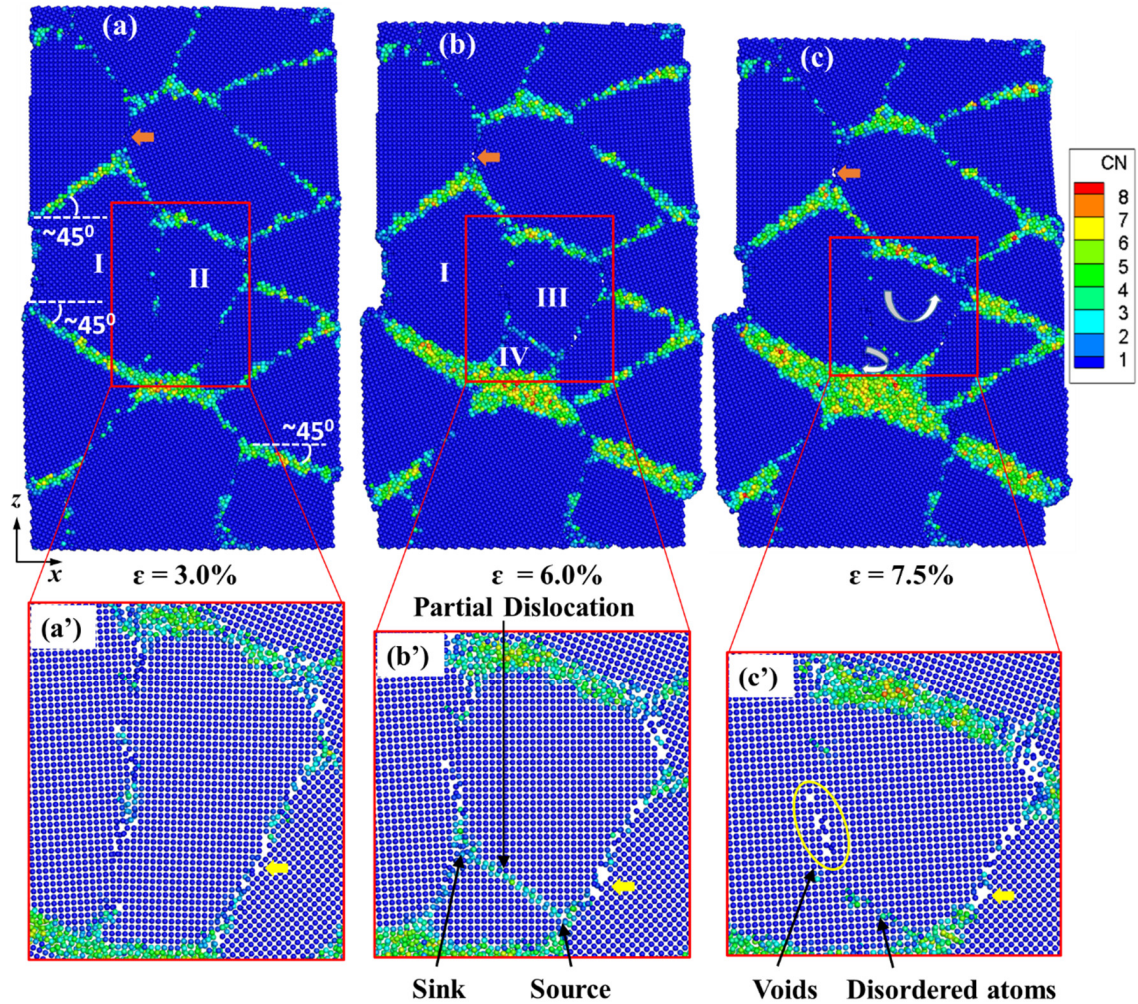

**Figure 2.** Side-view atomic configurations of 2D columnar polycrystalline YSTZ nanopillar at strains of (a) 3.0%, (b) 6.0% and (c) 7.5%. Atoms are colored according to CN. The orange arrows in (a-c) denote geometrical misfits of nanograins. The insets of (a'-c') illustrate the evolution of a partial dislocation and void formation at GBs. The yellow arrows and ellipse denote the growth and coalescence of nanovoids at GBs.

As plastic deformation takes place, in addition to the predominant deformation mechanisms of intergranular GB sliding and amorphous phase formation, two neighboring grains might rotate in a pattern that makes their orientation closer together. This type of GB rotation will lead to the elimination of the barrier presented by the GB between them. In the present study, we particularly consider the interface between grains I and II, as shown in Fig. 2a. Originally, the interface contains amorphous atoms and voids, but upon further loading, a partial dislocation nucleates from one GB, denoted as "source" in Fig. 2b'. This partial dislocation moves unimpeded across the whole grain II, with stacking faults left behind, until it encounters the opposing GB, which acts as a "sink", as shown in Fig. 2b'. The formation of partial dislocation splits grain II into two separated new grains: grain III and grain IV, as indicated in Fig. 2b, and then immediately rotation of these two grains occurs by disclination motion, as shown in Fig. 2c. Subsequently, most of the original amorphous atoms on the interfaces between grains I and II disappear, and meanwhile grains I, III, and IV merge to form a larger grain. This suggests that the generation of partial dislocation defect provides an alternative mechanism to GB sliding allowing rotation of nanograins during compression. This grain rotational accommodation mechanism is effective for superplastic deformation. Moreover, visual inspection of the atomic configurations reveals that nanovoids form due to this amorphous-to-crystal structure transition in between neighbor grains facilitated by grain rotation, as highlighted by the ellipse in Fig. 2c'. Previously, it was demonstrated by the atomic scale nanoindentation simulations [56] that dislocations can be emitted, absorbed, and even reflected from GBs. In our polycrystalline YSTZ nanopillar compression simulation, as the applied strain increases to 7.5%, the leading partial of this dislocation is absorbed by the GB but the trailing partial is not absorbed completely, which is referred to the disordered atoms left behind after the merging of grains in Fig. 2c'. The process of dislocation absorption is accompanied by rotation of grains, rearrangement of GB structure, and formation of nanovoids.

Consequently, in polycrystalline YSTZ nanopillars under uniaxial compressive loading, the cooperative intergranular deformation mechanisms of GB sliding, GB migration, grain rotation, and amorphous phase formation at GBs are observed. Also, the transgranular partial dislocations act as accommodations for overcoming the resistance against grain rotation; hence, the appearance of partial dislocations can be interpreted as a "grain redistribution" mechanism.

*3.2 Specimen size effect*

In this section, we study the specimen size effect on the mechanical properties of 3D polycrystalline YSTZ nanopillars. The obtained stress-strain responses for five nanopillars with different cross-sections (i.e., 10×10 nm$^2$, 15×15 nm$^2$, 20×20 nm$^2$, 25×25 nm$^2$, 30×30 nm$^2$) and the same height (40 nm) as described in Table 1, are plotted in Fig. 3. Despite some small differences, all these stress-strain curves show a very similar pattern; after the initial linear elastic region, a non-linear elastic response is observed, and then each nanopillar undergoes a relatively steady plastic deformation. The nonlinear elasticity is believed to be due to GB sliding. Compared with single crystalline YSTZ nanopillars [1], the yield stress of polycrystalline YSTZ nanopillars is significantly lower (almost three times lower). This observation again confirms the discussion above (in *section 3.1*) that GBs have a detrimental effect on the strength, which is consistent with the studies on other polycrystalline materials [57, 58]. The slope of the stress-strain curve in initial linear elastic region gives the Young's modulus ($E$) of the nanopillars. It can be noted from Fig. 3 that Young's modulus and yield stress (i.e., strength, $\sigma_y$) show a decreasing trend as the specimen size of polycrystalline YSTZ nanopillar is decreased, i.e., smaller nanopillars have lower strength. Such specimen size-induced softening is opposite to the usually observed "smaller is stronger" phenomenon in single crystal solids [20, 26, 27].

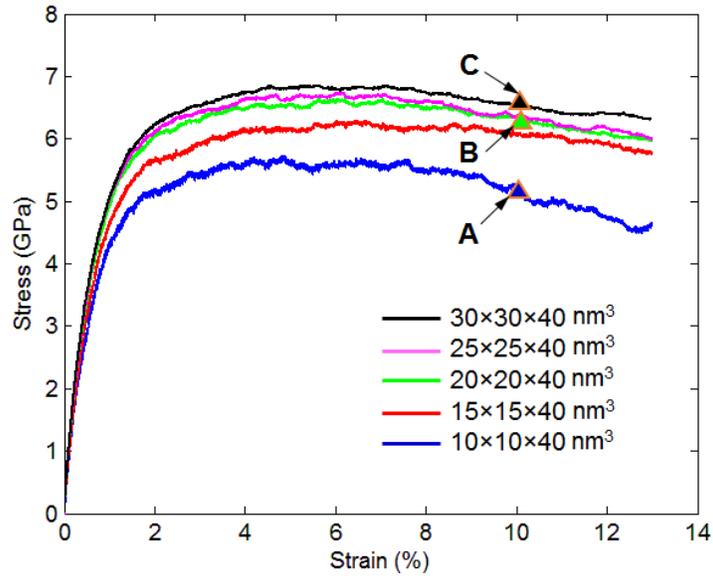

Figure 3. Stress-strain curves for polycrystalline YSTZ nanopillars as a function of specimen size.

To reveal the underlying softening mechanism of polycrystalline YSTZ nanopillars with decreasing cross section size, we track the deformation trajectories of various specimens and plot three representatives in Fig. 4. The deformed configurations in Fig. 4d-f correspond to the specific points, A, B and C, in Fig. 3, respectively. One can notice that in the nanopillar with size of 30 × 30 × 40 nm$^3$ (Fig. 4f), GB sliding only occurs at surface grains. In contrast, a plastic bending/buckling deformation behavior is observed in the smallest pillar (10 × 10 × 40 nm$^3$) (Fig. 4d), which has also been observed in the experimental investigation of nanocrystalline metal pillars [29].

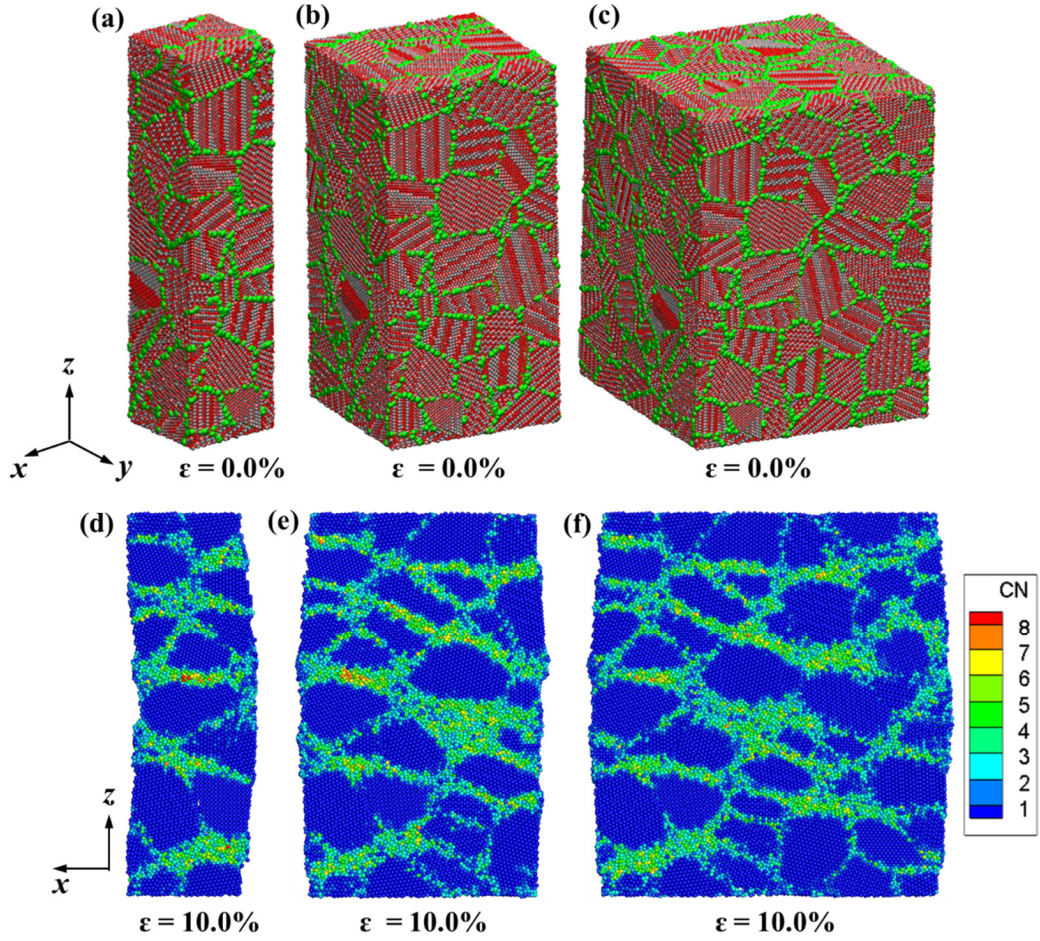

Figure 4. Representative models of polycrystalline YSTZ nanopillars with sizes of (a) 10 × 10 × 40 nm³, (b) 20 × 20 × 40 nm³ and (c) 30 × 30 × 40 nm³ for the specimen size effect study. The cross-sectional side views of the deformed atomic configurations in (a-c) at strain of 10.0%, corresponding to the points of A, B and C in Fig. 3, are shown in (d-f). Coordinate number (CN) is adopted to track the evolution of deformation. Blue color represents YSTZ with tetragonal structure, while green and other colors represent GBs and high strained atoms.

It is well known that free surfaces and GBs play important roles in mechanical response of polycrystalline nanopillars. However, since all the models in study of specimen size effect in this section have the same GB density, the impact from GBs may be negligible. In single crystalline solids, free surfaces serve as preferential nucleation sites for defects, however, in polycrystalline nanopillars they act as energy release exits through GB sliding; this is because the required energy for initiation of GB sliding is lower than that of surface defect. Therefore, we calculate the characteristic dimension ratio of free surface-to-volume ($A/V$) to quantitatively analyze the relation between free surface and mechanical properties of polycrystalline YSTZ nanopillars. Fig. 5 depicts the variations in free surface-to-volume ratio, Young's modulus and strength with respect to the specimen size; here we consider the width ($w$) of the cross section of polycrystalline YSTZ nanopillars as the specimen size indicator. It can be seen in Fig. 5b, c that the Young's modulus and strength decrease nonlinearly with decrease of the width of nanopillars. In contrast, $A/V$ shows a nonlinear increasing trend as the width is decreasing (Fig. 5a). These observed trends in mechanical properties and $A/V$, along with a close inspection of the polycrystalline YSTZ nanopillars deformation snapshots in Fig. 4d-f, suggest that with the increase in $A/V$, the travel path for GB sliding becomes shorter, in other words, the barrier against GB energy release is weakened, which directly leads to this specimen size-induced softening behavior (decrease in Young's modulus and strength). Since GB sliding prefers to occur at GBs that are near the free surfaces, nanopillars with larger $A/V$ yield at a lower stress simply because the fraction of grains that are at the free surfaces increases with increasing $A/V$. Therefore, it is not surprising to observe in Fig. 4d that with the decrease of nanopillar size, surface corrugations become easier to form.

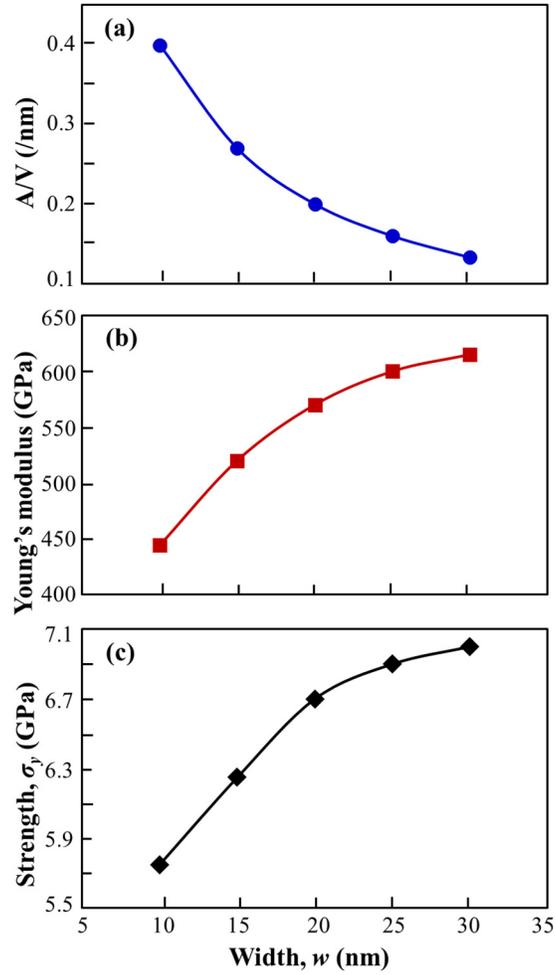

Figure 5. Correlations between free surface-to-volume ratio (*A/V*), mechanical properties and the specimen size of polycrystalline YSTZ nanopillars: (a) *A/V*, (b) Young's modulus, and (c) strength, as a function of width of the cross section.

## *3.3 Grain size effect*

In this section, we study the grain size effect on mechanical properties of polycrystalline YSTZ nanopillars. Fig. 6 shows the stress-strain curves of nanopillars with different grain sizes under compression. A noticeable decrease in Young's modulus and strength is observed as the grain size decreases. Such accelerated decrease in Young's modulus and strength indicate that in polycrystalline YSTZ nanopillars with an average grain size below ~15 nm, the mechanical properties are very sensitive to the average grain size.

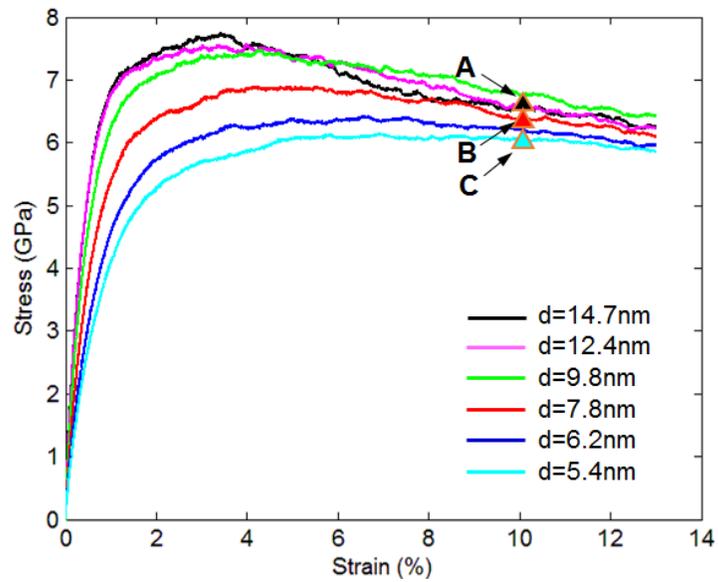

Figure 6. Stress-strain curves for polycrystalline YSTZ nanopillars as a function of average grain size.

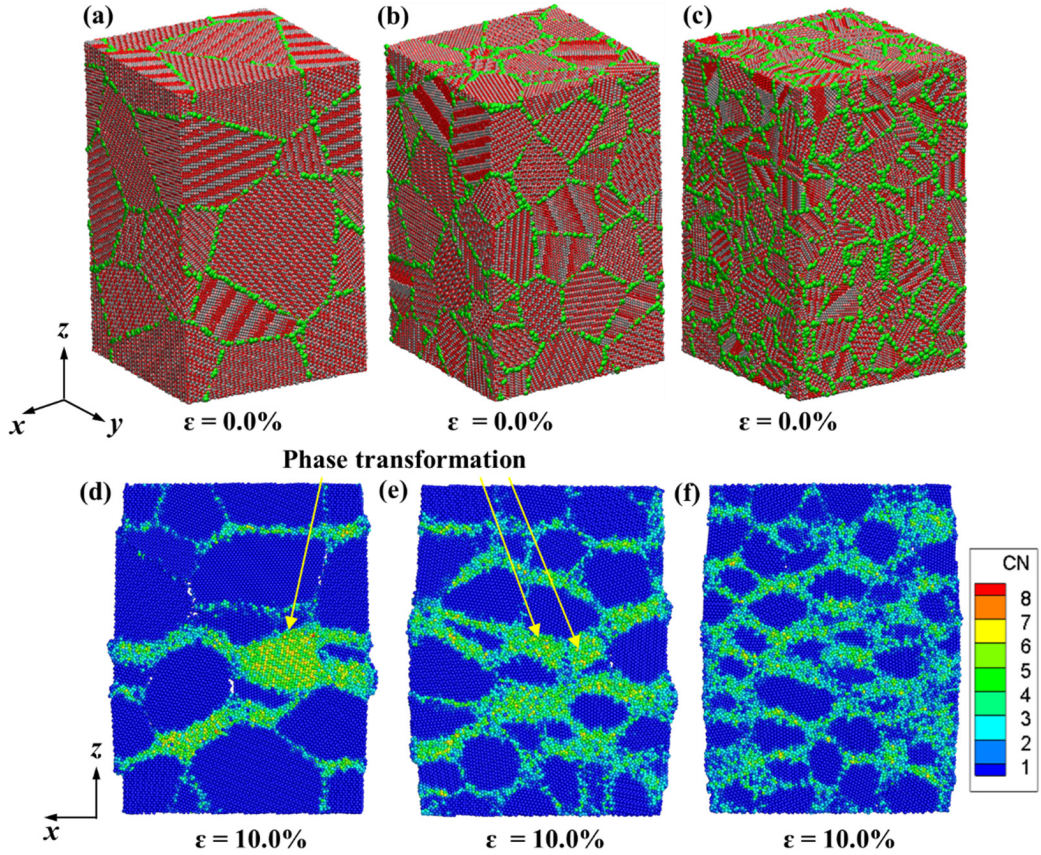

Figure 7. Representative polycrystalline YSTZ nanopillars with average grain sizes of (a) 14.7 nm, (b) 7.8 nm and (c) 5.4 nm for grain size effect study. (d-f) show the corresponding cross-sectional side views of atomic configurations in (a-c) at strain of 10.0%, i.e., the points of A, B and C in Fig. 6. Coordinate number (CN) is adopted to track the evolution of deformation. Blue color represents YSTZ with tetragonal structure, while green and other colors represent GBs and high strained atoms.

In order to better understand the deformation process, we present the atomic configurations of three representative nanopillars with average grain sizes of 14.7 nm, 7.8 nm and 5.4 nm at a strain of 10.0% in Fig. 7d-f, respectively, corresponding to the specific points of A, B and C in Fig. 6. It should be mentioned that in all these nanopillars, intergranular deformation occurs before transgranular deformation, as it is a predominant deformation mechanism due to the role that GBs play. In other words, the decrease in Young's modulus should be attributed to the release of stored elastic energy by local plastic deformation at GB regions. As the grain size decreases, the resistance to intergranular deformation becomes even smaller, especially amorphous phase formation on GBs, becomes smaller, which can be confirmed by the homogenous deformation in 5.4 nm grains nanopillar (Fig. 7f). In contrast, the dominant intergranular deformation, mainly GB sliding, in 14.7 nm grains nanopillar encounters large resistance, and alternatively transgranular deformation of phase transformation, as shown in Fig. 7d, is stimulated. Since GB sliding occurs in-between the grains near the free surface, a transition of intergranular deformation from amorphous phase formation (Fig. 7f) to GB sliding (Fig. 7d) takes place with the increase of grain size. This can be verified by the deformed atomic configurations in Fig. 7d-f that clear surface steps left behind after compression, and this is due to GB sliding in the nanopillar with the average grain size of 14.7 nm (Fig. 7d), but no obvious surface steps are observed in the nanopillar with the average grain size of 5.4 nm (Fig. 7f). Meanwhile in the nanopillar with the average grain size of 14.7 nm, the geometrical misfits lead to the formation of voids and cracks on GBs. In the nanopillar with the average grain size of 7.8 nm (Fig. 7e), both GB deformation and transgranular phase transformation are observed. It is well known that the crystalline grains are much harder than the GBs, therefore the competition between intergranular GB sliding and transgranular deformation, i.e., phase transformation, in nanopillars with large grains is responsible for the enhancement of Young's modulus and strength.

As discussed above, GBs are weak links in polycrystalline YSTZ nanopillars, and they serve as sliding path and nucleation sites of defects (e.g., amorphous phase, dislocation, voids, and cracks). Therefore, for the nanopillars with the same specimen size and different grain sizes, it is useful to calculate the fraction of GBs in the structure, i.e., GB density ($\rho$), and study its correlation to the mechanical properties. Fig. 8a displays a nonlinear relation between GB density and the average grain size. It can be seen that with the decrease of grain size, GB density increases exponentially. Such nonlinear increasing trend in GB density can be utilized to interpret the nonlinear decreasing trends in Young's modulus (Fig. 8b) and strength (Fig. 8c), where a smaller grain size implies a larger GB density and therefore lowers the nanopillar strength.

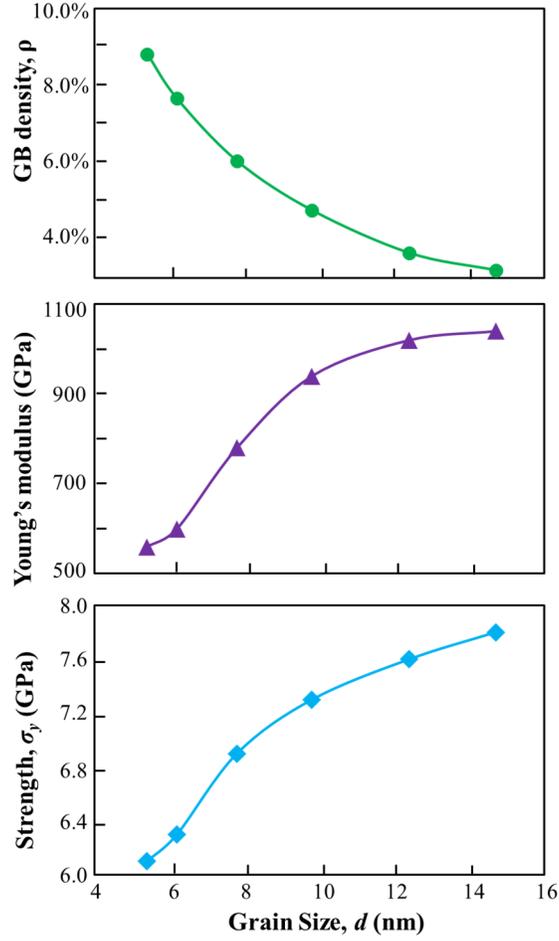

Figure 8. Correlations between GB density, mechanical properties and the average grain size of polycrystalline YSTZ nanopillars: (a) GB density, (b) Young's modulus, and (c) strength, as a function of the grain size.

The grain size-induced softening phenomenon that observed in the current work is very similar to that revealed in the nanocrystalline metallic materials [24, 28, 30, 59]. The classical Hall-Petch relation for bulk metals is:

$$\sigma_y = \sigma_0 + kd^2, \tag{1}$$

where $\sigma_0$ is the friction stress in the absence of GBs, $k$ is a constant and $d$ is the average grain size. This relation describes the characteristic increase in strength of polycrystalline metals with decreasing grain size, down to ~40 nm. When the grain size decreases to below 20 nm, it is revealed that the slope of Hall-Petch relation (i.e., $k$ value in Eq. (1)) becomes negative, which is known to be as the "inverse" Hall-Petch relation [24]. To quantitatively address the relationship between strength and average grain size, we plot the relationship between the strength and the inverse square root of grain size $d^{-1/2}$ in Fig. 9 for polycrystalline YSTZ nanopillars with specimen size of 25.0 × 25.0 × 40.0 nm³ (Model I) and 20.0 × 20.0 × 40.0 nm³ (Model II). As it can be clearly seen, the strength of nanopillar decreases linearly with $d^{-1/2}$ in the range of 0.26 and 0.43, corresponding to the average grain size between 14.7 and 5.4 nm. Inverse Hall-Petch equations for Model I (Eq. (2)) and Model II (Eq. (3)) are obtained, respectively:

$$\sigma_y = 10.557 - 10.388 d^{-1/2}, \tag{2}$$

$$\sigma_y = 10.253 - 10.111 d^{-1/2}. \tag{3}$$

It can be noted that the obtained two curves in Fig. 9 are almost parallel to each other, except that the strength of Model I are higher than its correspondence in Model II, which is consistent to the conclusion in section 3.2 regarding the specimen size effect. The calculated friction stresses of 10.557 and 10.253 GPa and slopes of -10.388 and -10.111 Gpa/nm$^{-1/2}$ for Model I and Model II, respectively, are in good agreements with the values reported for ceramics [59].

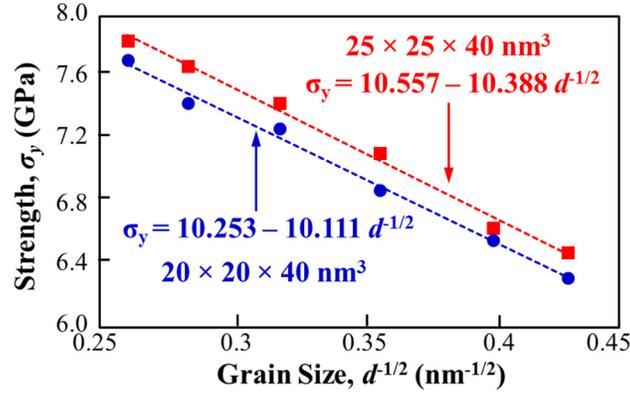

Figure 9. Strength of polycrystalline YSTZ nanopillars with two different specimen sizes as a function of the inverse square root of grain size, $d^{-1/2}$. Inverse linear Hall-Petch relations can be observed.

## 4. Conclusions

The intergranular and transgranular plastic deformation mechanisms, as well as the effects of specimen- and grain-size on the Young's modulus and strength of polycrystalline YSTZ nanopillars were investigated using molecular dynamics simulations. Cooperative intergranular deformations consisting of local GB sliding, GB migration, amorphous phase formation at GBs, and grain rotation are dominant deformation mechanisms for YSTZ nanopillars under compression. Nucleation and propagation of partial dislocations were also observed to split the large grains, which furthermore accommodated grain rotations. Formation of voids due to amorphous-to-crystalline phase transition between neighboring grains, as well as cracks and voids induced by geometrical misfits were observed.

The Young's modulus and strength of the YSTZ nanopillars were found to decrease with the decreasing of specimen size. Since the dominant mechanism of GB sliding prefers to occur in grains near the free surfaces, the free surface-to-volume ratio was calculated and it was found to be responsible for the observed specimen size-induced softening phenomenon. The experimentally observed bending/buckling of nanopillars was also manifested in our simulation of the smallest polycrystalline YSTZ nanopillar (with the cross-section width of 10 nm).

Grain size also shows a strong influence on the mechanical properties of polycrystalline YSTZ nanopillars. The values of Young's modulus and strength are dominated by the density of GBs. A transition from amorphous phase formation on GBs to a competition between GB sliding and transgranular phase transformation was observed, and it is responsible for the increase of strength in nanopillar with large grain size. Furthermore, inverse Hall-Petch equations were constructed in this grain size-induced softening region: 5 nm $< d <$ 15 nm. Our current work presents a comprehensive analysis on the fundamental nanoscale deformation mechanisms, as well as the specimen- and grain-size dependence on mechanical properties of polycrystalline YSTZ nanopillars. This work provides some necessary information for the development of nanocrystalline zirconia pillars and other potential shape memory ceramics. Further studies are required to investigate the elemental segregation at GBs and their effects on the structural characteristics of GBs and the consequential effects on mechanical properties and deformation mechanisms of nanoscale ceramics.

**Acknowledgement**

This work was supported by the U.S. Department of Energy, Office of Science, Basic Energy Sciences, under Award number DE-SC0019279. The authors are grateful for computer time allocation provided by the Extreme Science and Engineering Discovery Environment (XSEDE).